\documentclass[5p]{elsarticle}

\usepackage{graphics}
\usepackage{graphicx}
\usepackage{epsfig}

\usepackage{amssymb}
\journal{TIPP09 Proceedings in NIMA}

\begin{document}

\begin{frontmatter}

\title{Photon generation by laser-Compton scattering at the KEK-ATF}

\author[hiroshima]{S.~Miyoshi}
\author[hiroshima]{T.~Akagi}
\author[kek]{S.~Araki}
\author[kek]{Y.~Funahashi}
\author[waseda]{T.~Hirose}
\author[kek]{Y.~Honda}
\author[hiroshima]{M.~Kuriki}
\author[ihep]{X.~Li}
\author[kek]{T.~Okugi}
\author[kek]{T.~Omori}
\author[ihep]{G.~Pei}
\author[waseda]{K.~Sakaue}
\author[kek]{H.~Shimizu}
\author[hiroshima]{T.~Takahashi}
\author[kek]{N.~Terunuma}
\author[kek]{J.~Urakawa}
\author[hiroshima]{Y.~Ushio}
\author[waseda]{M.~Washio}

\address[hiroshima]{Graduate School of Advanced Sciences of Matter, Hiroshima University, Higashi-Hiroshima, Hiroshima 739-8530, Japan}
\address[kek]{High Energy Accelerator Research Organization, Tsukuba, Ibaraki 305-0801, Japan}
\address[waseda]{Research Institute for Science and Engineering, Waseda University, Tokyo 162-0044, Japan}
\address[ihep]{Institute of High Energy Physics, Chinese Academy of Sciences, Beijing 100049, China}

\begin{abstract}

We performed a photon generation experiment by laser-Compton scattering at the KEK-ATF, 
aiming to develop a Compton based polarized positron source for linear colliders. 
In the experiment, laser pulses with a 357~MHz repetition rate were accumulated 
and their power was enhanced by up to 250 times in the Fabry-Perot optical resonant cavity. 
We succeeded in synchronizing the laser pulses and colliding them 
with the 1.3~GeV electron beam in the ATF ring while maintaining 
the laser pulse accumulation in the cavity. 
As a result, we observed 26.0$\pm$0.1 photons per electron-laser pulse crossing, 
which corresponds to a yield of 10$^8$ photons in a second. 
\end{abstract}


\end{frontmatter}

\section{Introduction}
\label{intro}
The International Linear Collider (ILC) seeks to provide polarized electron beams 
for precise measurements of physics phenomena. 
In addition to the electron beam, 
there are hopes that polarizing positron beams 
as well will further improve the performance of the ILC as a machine 
for making precise measurements \cite{roleofposipol}. 
The polarized positrons are generated via pair creation 
by impinging polarized photons of about 10~MeV into target materials. 
Therefore, the generation of polarized intense photons 
of the 10~MeV energy regions would be a key technology for the polarized positron source. 
The ILC baseline design adopts the helical undulator scheme 
in which 150~GeV electrons are fed into the helical undulator 
with a length of more than 150~m to create photons \cite{rdr}. 
Though the undulator scheme is the baseline design, 
a 150~m-long undulator has yet to be developed 
and we will have to wait until the construction of the main LINAC 
of the ILC before an electron beam of 150~GeV can be provided. 

One option for polarized positron generation is to use laser-Compton scattering 
for photon generation (the Compton scheme). 
The advantage of the Compton scheme is that the required energy of the electron beam 
to generate 10~MeV photons is about 1~GeV, 
which is low enough to develop positron sources 
using the existing small electron beam facilities. 
In addition to the energy of the electron beam, 
the polarization of the generated positrons can be easily controlled 
in the Compton scheme by changing the polarization of the laser, 
though accomplishing this with the undulator scheme is neither a self-evident nor easy task. 
Thus the Compton scheme is being developed in parallel with the undulator scheme. 
We report in this paper on the status of the R\&D efforts 
for the Compton scheme in the KEK-ATF. 

\begin{figure}[htb]
\begin{center}
\includegraphics[width=7.1cm,clip]{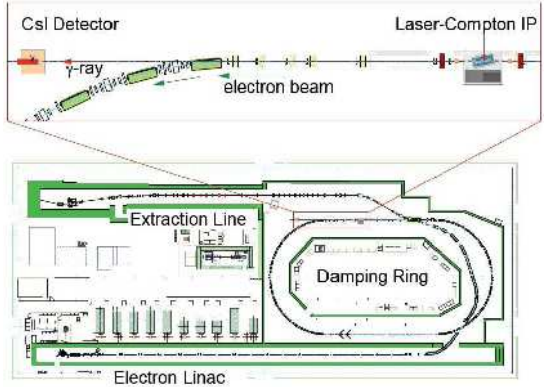}
\caption{(color) 
The optical resonant cavity is installed in the straight section of the KEK-ATF damping ring \cite{shimizu}. 
A CsI detector for photon detection is placed the 18 m downstream of the laser electron interaction point in the optical resonant cavity. }
\label{fig:atf}
\end{center}
\end{figure}

The proof of the principle of generating polarized positrons 
with the Compton scheme has been demonstrated 
by a series of experiments performed at the ATF \cite{atfposipol}. 
The next step toward the positron source will be to increase the intensity 
of photons required by the ILC. 
For this purpose, we adopt a scheme to increase the intensity 
of the laser pulses by accumulating them in an optical resonant cavity. 
Since accumulation of laser pulses in the cavity requires precise control 
of the optical system, 
and the ATF requires sophisticated machine operation 
to maintain the ultra-low emittance of its beam, 
it is no easy task to accommodate an optical system 
in the accelerator environment and to make laser pulses collide with the electron beam. 
Therefore we installed a Fabry-Perot type optical resonant cavity in the ATF damping ring, 
as shown in Figure~\ref{fig:atf}, and performed an experiment to generate photons by laser-Compton scattering.

\section{Experimental Setup}
\label{setup}

The beam energy and typical beam size 
of the ATF are 1.3~GeV and 100~$\mu$m (horizontal) $\times$ 10~$\mu$m (vertical) in the root mean square, respectively. 
The repetition rate of laser pulses is 357~MHz, 
which is the same as the electron bunch spacing in the ATF. 
The wavelength and pulse width of the laser are 1064~nm and 5~ps in the root mean squire. 
The crossing angle of the laser pulse and the electron beam is 12~degrees, 
which determined the maximum energy of the Compton gamma at 28~MeV. 
The energy of the photon was discriminated by the slit placed between the laser-electron interaction point 
and the CsI photon detector. 
As a result, the energy of photons detected in the detector is 19~MeV to 28~MeV with the average of 24~MeV.

\begin{figure}[htb]
\begin{center}
\includegraphics[width=4.9cm,clip]{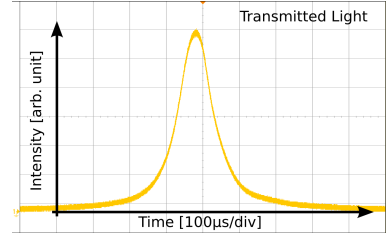}
\caption{(color) The intensity of the transmitted light from the optical resonant cavity. 
The yellow line is the transmitted laser power. 
The peak for intensity shows the cavity is on-resonance. }
\label{fig:scan}
\end{center}
\end{figure}

In order to accumulate laser pulses in the optical resonant cavity, 
the cavity has to be on-resonance with the injected laser pulses. 
It requires the length of the optical resonant cavity to be integer times the half wave length 
of the laser wave precisely. 
In addition, accumulation of mode locked pulsed lasers requires that the length 
of the optical resonant cavity is integer times those of the cavity inside the laser oscillator. 
Therefore the length of the optical resonant cavity is 420 mm. 
The waist size of laser pulses is 30 $\mu$m (1$\sigma$) in the optical resonant cavity, 
because the curvature radius of mirrors is 210.5 mm. 
The resonant condition of the optical resonant cavity is monitored by the intensity 
of the transmitted light from the cavity, which is at the maximum when the cavity is on-resonance. 
The intensity of the transmitted light obtained by changing the length of the cavity is shown in Figure~\ref{fig:scan}. 
We observed a resonant peak with a width of 0.7~nm which indicated 
that the length of the optical resonant cavity has to be controlled within this precision. 
The resonant condition in term of the finesse was estimated from the width and was about 1000, 
which is consistent with expectations from the mirror reflectivity of 99.6~\%, within the error. 

\begin{figure}[htb]
\begin{center}
\includegraphics[width=8.6cm,clip]{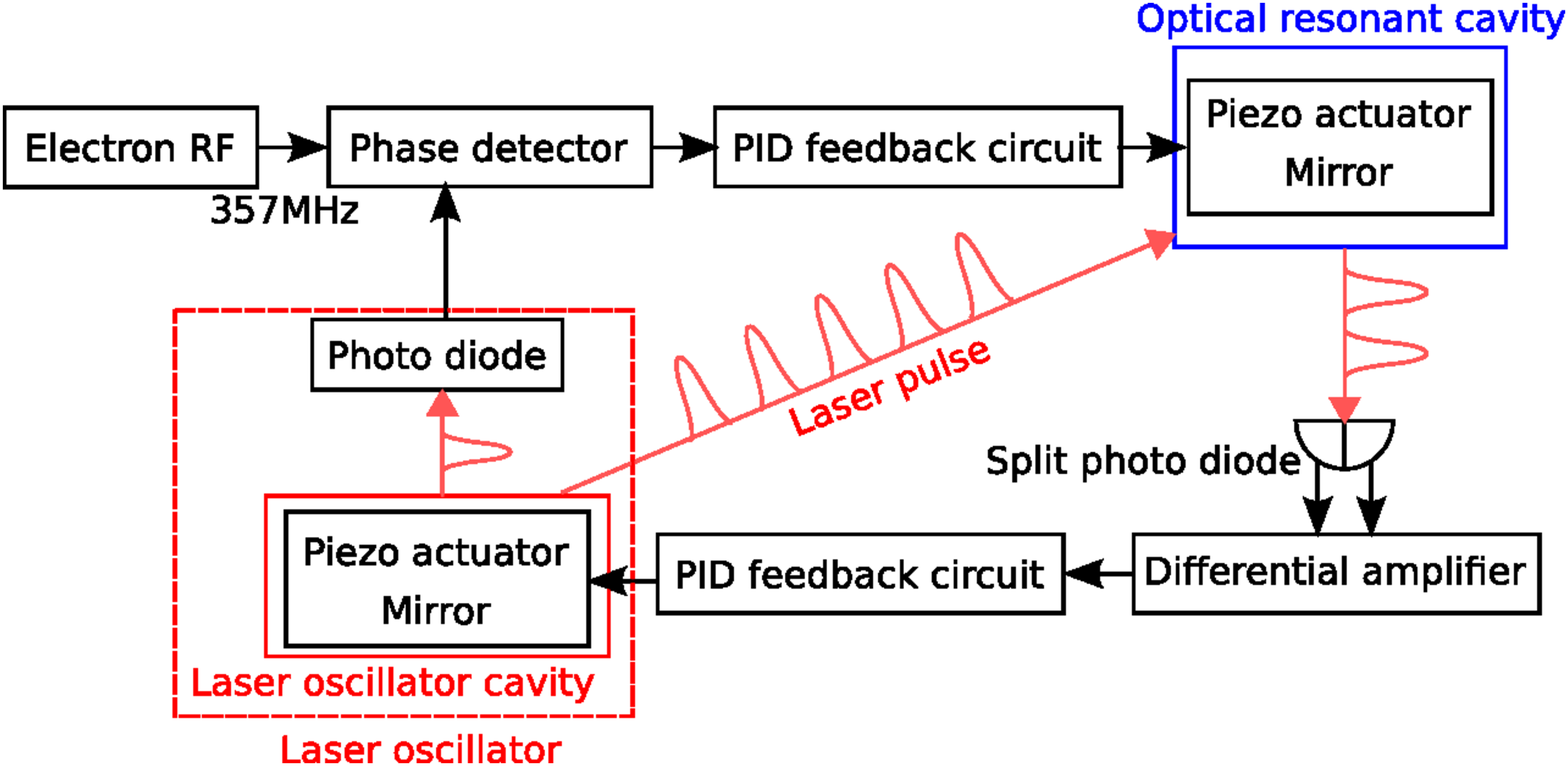}
\caption{Schematic of the feedback system for the laser pulse accumulation and timing synchronization. 
See text and \cite{tiltlock} for detail. }
\label{fig:fbsys}
\end{center}
\end{figure}

\begin{figure}[htb]
\begin{center}
\includegraphics[width=9.4cm,clip]{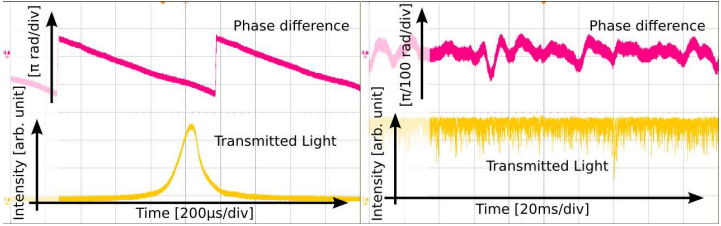}
\caption{(color) The intensity of the transmitted light (yellow) , 
and the relative phase between ATF master signal and laser pulse (red) when the feedback is turned off (left) 
and on(right). 
The intensity of the transmitted light in the right figure is constant and its value corresponds 
to the peak in the left figure. 
The phase signal in the right hand figure is also constant with some fluctuation of about 5~ps. }
\label{fig:resonance}
\end{center}
\end{figure}

To accumulate laser pulses in the optical resonant cavity and continuously make collisions 
of laser pulses and electron bunches, 
the length of the optical cavity has to be kept within the nm range 
and the timing between laser pulses and electron bunches has to be synchronized within ps precision. 
For this purpose we developed the feedback system as schematically shown in Figure~\ref{fig:fbsys}, 
in which special care was taken to achieve stable resonant and timing synchronization \cite{sakaue}. 
Firstly, the feedback system, consisting of a mode locked laser and the resonant cavity, 
forms a single closed loop for stable feedback operation. 
Secondly, the system was constructed taking the required speed for the feedback operation into account. 
Keeping the resonance of the resonant cavity needs relatively faster feedback 
while the timing synchronization can be relatively slow. 
On the other hand, the length of the optical cavity cannot be changed rapidly 
as it uses relatively large mirrors while laser wave length can change faster as it uses smaller size mirrors. 
Therefore, the resonant signal from the optical cavity is fed back to the mode locked laser 
while the timing synchronization is achieved by adjusting the length of the optical cavity. 
The performance of the feedback system is shown in Figure~\ref{fig:resonance}. 
The left hand of Figure~\ref{fig:resonance} shows the transmitted laser power 
while sweeping the length of the optical resonant cavity. 
The right hand of Figure~\ref{fig:resonance} shows the resonance and synchronization signal 
with the feedback loop system turned on. 
It shows that we successfully achieved stable resonance of the cavity with the laser wave 
as well as timing synchronization of the laser pulses and the electron bunches.

\begin{table*}[htb]
\caption{Summary of experimental results and conditions. 
There are additional 3~\% systematic errors common for all ``Gamma'' measurements. 
There are additional 14~\% systematic errors common for all ``Laser power'' measurements. 
}
\label{table2}
\begin{center}
\begin{tabular}{|c||c|c|c|c|c|}
\hline
\# of bunches & e$^-$ current[mA] & Laser power[W] & Gamma[1/train] & Simulation & Normalize[gamma/A$\cdot$W]\\
\hline
1 & 2.2 & 437$\pm$2 & 5.4$\pm$0.3 & 4.9$\pm$0.3 & 5.6$\pm$0.3 \\ 
5 & 4.7 & 423$\pm$2 & 10.6$\pm$0.1 & 10.5$\pm$0.5 & 5.3$\pm$0.1\\
10 & 8.5 & 470$\pm$2 & 19.0$\pm$0.1 & 21$\pm$1 & 4.8$\pm$0.1\\
15 & 11 & 498$\pm$2 & 26.0$\pm$0.1 & 29$\pm$1 & 4.8$\pm$0.1\\
\hline
\end{tabular}
\end{center}
\end{table*}

\section{Experiment}
\label{exp}

\begin{figure}[h]
\begin{center}
\includegraphics[width=5.7cm,clip]{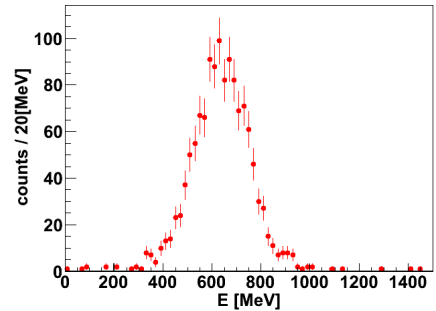}
\caption{Energy deposit in the CsI detector observed from laser-electron interaction with 15 electron bunches/train. }
\label{fig:deposit}
\end{center}
\end{figure}

We performed detection of Compton photons with 1,5,10 and 15 electron bunches/train in the ATF. 
First, we scanned the position of the optical resonant cavity relative 
to the electron beam to find the optimum position. 
Then, the timing of the laser pulses relative to the electron bunches was adjusted 
by changing relative phase between the ATF master oscillator and laser pulses. 
Figure~\ref{fig:deposit} shows the energy deposit of Compton photons in the CsI detector 
which was placed 18~m downstream of the laser-electron interaction point. 
At 15-bunch operation, we observed 26.0$\pm$0.1 photons/train. 
Since the revolution rate of the electron bunches in the ATF is 2.16~MHz, 
it can be estimated that 10$^8$ photons are generated per second in all solid angles. 

The results of the experiment under each condition, 
i.e. the number of bunches per train and the laser power, 
are summarized in Table~\ref{table2}. 
The expected yields, shown as ``Simulation'' in the table, were calculated by the simulation code CAIN \cite{cain}. 
The ambiguity of the simulation arose mainly from the uncertainty 
of the electron bunch length during the experiment, 
which varied from 6.0~mm to 9.0~mm depending on accelerator conditions. 
The photon yields were consistent with the simulation with a smaller number of bunches per train, 
while they deviated from the simulation with a larger number of bunches per train. 
The reason for these deviations is under investigation. 
We suspect that it may be due to synchrotron oscillation of electron bunches, 
as this effectively enlarges the average bunch length in a train. 
The bunch oscillation was investigated using a streak camera and larger oscillation 
for larger number of bunches was observed. 
However, a detailed comparison with the photon yield taken under accelerator conditions is still in progress. 

\section{Conclusions}
\label{conclu}
The optical resonant cavity was constructed and installed in the KEK-ATF electron ring 
for the purpose of the photon generation by laser-Compton scattering, 
aimed at the development of a polarized positron source. 
We succeeded in accumulating laser pulses in the optical resonant cavity 
and synchronizing them with the electron bunches simultaneously under the operation of the ATF. 
The laser power was enhanced up to 250 times that of injection power by the accumulation. 
The generated photons were 26.0$\pm$0.1 photons/train, 
which corresponds to 10$^8$/second. 
The results demonstrated the feasibility of using the optical resonant cavity 
for effective photon generation by laser-Compton scattering. 
The photon yields were almost consistent with expectations 
but appreciable deviation was observed for larger number of bunches per train. 
The effect of synchrotron oscillation on the bunch in the ATF may be a cause 
but further investigation is necessary.

\section*{Acknowledgements}
The authors would like to thank ATF collaborators for discussing and helping. 
This work was supported in part by 
KAKENHI [(B)18340076, 17GS0210], 
Quantum Beam Technology Program by JST, 
the KEK Promotion of collaborative research programs in universities 
and the Hiruma research funds at Hiroshima University.


\begin{thebibliography}{00}
\bibitem{roleofposipol}G.~Moortgat-Pick {\it et} al., Physics Reports, {\bf 460} (2008) 131
\bibitem{rdr}ILC-REPORT-2007-001(2007)
\bibitem{atfposipol}T.~Omori {\it et} al., Phys. Rev. Letts {\bf 96} (2006) 114801
\bibitem{shimizu}H.~Shimizu {\it et} al., J. Phys. Soc. Jpn. {\bf 78} (2009) 074501
\bibitem{sakaue}K.~Sakaue, Doctoral thesis, Waseda University (2009)
\bibitem{tiltlock}D.A.~Shaddock {\it et} al., Opt. Lett. {\bf 24} (1999) 1499
\bibitem{cain}K.~Yokoya, http://lcdev.kek.jp/yokoya/CAIN/cain235/
\end{thebibliography}
\end{document}